\documentclass{llncs}
\usepackage{llncsdoc}
\usepackage{amssymb,amsmath,psfrag}
\usepackage{color,graphicx}
\graphicspath{{/mnt/partage/New/BSS_Image/Bss/Bss_1D/Eps2/},{/mnt/partage/New/Hermes/pbivision/Introduction/Eps2/}}

\def\bdoc{\begin{document}}             \def\edoc{\end{document}}
\def\babs{\begin{abstract}}             \def\eabs{\end{abstract}}
\def\bcc{\begin{center}}                \def\ecc{\end{center}}
\def\ben{\begin{enumerate}}             \def\een{\end{enumerate}}
\def\bit{\begin{itemize}}               \def\eit{\end{itemize}}
\def\beq{\begin{equation}}              \def\eeq{\end{equation}}
\def\btabu{\begin{tabular}}             \def\etabu{\end{tabular}}
\def\barr{\begin{array}}                \def\earr{\end{array}}
\def\beqn{\begin{eqnarray}}             \def\eeqn{\end{eqnarray}}
\def\beqnx{\begin{eqnarray*}}           \def\eeqnx{\end{eqnarray*}}
\def\bfig{\begin{figure}}               \def\efig{\end{figure}}
\def\bpic{\begin{picture}}              \def\epic{\end{picture}}
\def\bm#1{\mbox{\boldmath $#1$}}
\def\wt#1{\widetilde{#1}}
\def\wh#1{\widehat{#1}}

\def\ab{{\bm a}}\def\Ab{{\bm A}}\def\Ac{{\cal A}}
\def\bb{{\bm b}}\def\Bb{{\bm B}}\def\Bc{{\cal B}}
\def\cb{{\bm c}}\def\Cb{{\bm C}}\def\Cc{{\cal C}}
\def\db{{\bm d}}\def\Db{{\bm D}}\def\Dc{{\cal D}}
\def\eb{{\bm e}}\def\Eb{{\bm E}}\def\Ec{{\cal E}}
\def\fb{{\bm f}}\def\Fb{{\bm F}}\def\Fc{{\cal F}}
\def\gb{{\bm g}}\def\Gb{{\bm G}}\def\Gc{{\cal G}}
\def\hb{{\bm h}}\def\Hb{{\bm H}}\def\Hc{{\cal H}}
\def\ib{{\bm i}}\def\Ib{{\bm I}}\def\Ic{{\cal I}}
\def\jb{{\bm j}}\def\Jb{{\bm J}}\def\Jc{{\cal J}}
\def\kb{{\bm k}}\def\Kb{{\bm K}}\def\Kc{{\cal K}}
\def\lb{{\bm l}}\def\Lb{{\bm L}}\def\Lc{{\cal L}}
\def\mb{{\bm m}}\def\Mb{{\bm M}}\def\Mc{{\cal M}}
\def\nb{{\bm n}}\def\Nb{{\bm N}}\def\Nc{{\cal N}}
\def\ob{{\bm o}}\def\Ob{{\bm O}}\def\Oc{{\cal O}}
\def\pb{{\bm p}}\def\Pb{{\bm P}}\def\Pc{{\cal P}}
\def\qb{{\bm q}}\def\Qb{{\bm Q}}\def\Qc{{\cal Q}}
\def\rb{{\bm r}}\def\Rb{{\bm R}}\def\Rc{{\cal R}}
\def\sb{{\bm s}}\def\Sb{{\bm S}}\def\Sc{{\cal S}}
\def\tb{{\bm t}}\def\Tb{{\bm T}}\def\Tc{{\cal T}}
\def\ub{{\bm u}}\def\Ub{{\bm U}}\def\Uc{{\cal U}}
\def\vb{{\bm v}}\def\Vb{{\bm V}}\def\Vc{{\cal V}}
\def\wb{{\bm w}}\def\Wb{{\bm W}}\def\Wc{{\cal W}}
\def\xb{{\bm x}}\def\Xb{{\bm X}}\def\Xc{{\cal X}}
\def\yb{{\bm y}}\def\Yb{{\bm Y}}\def\Yc{{\cal Y}}
\def\zb{{\bm z}}\def\Zb{{\bm Z}}\def\Zc{{\cal Z}}
\def\abh{\wh{\ab}} \def\Abh{\wh{\Ab}}
\def\bbh{\wh{\bb}} \def\Bbh{\wh{\Bb}}
\def\cbh{\wh{\cb}} \def\Cbh{\wh{\Cb}}
\def\dbh{\wh{\db}} \def\Dbh{\wh{\Db}}
\def\ebh{\wh{\eb}} \def\Ebh{\wh{\Eb}}
\def\fbh{\wh{\fb}} \def\Fbh{\wh{\Fb}}
\def\gbh{\wh{\gb}} \def\Gbh{\wh{\Gb}}
\def\hbh{\wh{\hb}} \def\Hbh{\wh{\Hb}}
\def\ibh{\wh{\ib}} \def\Ibh{\wh{\Ib}}
\def\jbh{\wh{\jb}} \def\Jbh{\wh{\Jb}}
\def\kbh{\wh{\kb}} \def\Kbh{\wh{\Kb}}
\def\lbh{\wh{\lb}} \def\Lbh{\wh{\Lb}}
\def\mbh{\wh{\mb}} \def\Mbh{\wh{\Mb}}
\def\nbh{\wh{\nb}} \def\Nbh{\wh{\Nb}}
\def\obh{\wh{\ob}} \def\Obh{\wh{\Ob}}
\def\pbh{\wh{\pb}} \def\Pbh{\wh{\Pb}}
\def\qbh{\wh{\qb}} \def\Qbh{\wh{\Qb}}
\def\rbh{\wh{\rb}} \def\Rbh{\wh{\Rb}}
\def\sbh{\wh{\sb}} \def\Sbh{\wh{\Sb}}
\def\tbh{\wh{\tb}} \def\Tbh{\wh{\Tb}}
\def\ubh{\wh{\ub}} \def\Ubh{\wh{\Ub}}
\def\vbh{\wh{\vb}} \def\Vbh{\wh{\Vb}}
\def\wbh{\wh{\wb}} \def\Wbh{\wh{\Wb}}
\def\xbh{\wh{\xb}} \def\Xbh{\wh{\Xb}}
\def\ybh{\wh{\yb}} \def\Ybh{\wh{\Yb}}
\def\zbh{\wh{\zb}} \def\Zbh{\wh{\Zb}}
\def\abt{\wt{\ab}} \def\Abt{\wt{\Ab}}
\def\bbt{\wt{\bb}} \def\Bbt{\wt{\Bb}}
\def\cbt{\wt{\cb}} \def\Cbt{\wt{\Cb}}
\def\dbt{\wt{\db}} \def\Dbt{\wt{\Db}}
\def\ebt{\wt{\eb}} \def\Ebt{\wt{\Eb}}
\def\fbt{\wt{\fb}} \def\Fbt{\wt{\Fb}}
\def\gbt{\wt{\gb}} \def\Gbt{\wt{\Gb}}
\def\hbt{\wt{\hb}} \def\Hbt{\wt{\Hb}}
\def\ibt{\wt{\ib}} \def\Ibt{\wt{\Ib}}
\def\jbt{\wt{\jb}} \def\Jbt{\wt{\Jb}}
\def\kbt{\wt{\kb}} \def\Kbt{\wt{\Kb}}
\def\lbt{\wt{\lb}} \def\Lbt{\wt{\Lb}}
\def\mbt{\wt{\mb}} \def\Mbt{\wt{\Mb}}
\def\nbt{\wt{\nb}} \def\Nbt{\wt{\Nb}}
\def\obt{\wt{\ob}} \def\Obt{\wt{\Ob}}
\def\pbt{\wt{\pb}} \def\Pbt{\wt{\Pb}}
\def\qbt{\wt{\qb}} \def\Qbt{\wt{\Qb}}
\def\rbt{\wt{\rb}} \def\Rbt{\wt{\Rb}}
\def\sbt{\wt{\sb}} \def\Sbt{\wt{\Sb}}
\def\tbt{\wt{\tb}} \def\Tbt{\wt{\Tb}}
\def\ubt{\wt{\ub}} \def\Ubt{\wt{\Ub}}
\def\vbt{\wt{\vb}} \def\Vbt{\wt{\Vb}}
\def\wbt{\wt{\wb}} \def\Wbt{\wt{\Wb}}
\def\xbt{\wt{\xb}} \def\Xbt{\wt{\Xb}}
\def\ybt{\wt{\yb}} \def\Ybt{\wt{\Yb}}
\def\zbt{\wt{\zb}} \def\Zbt{\wt{\Zb}}
\def\alphab{\bm{\alpha}}
\def\betab{\bm{\beta}}
\def\gammab{\bm{\gamma}}            \def\Gammab{\bm{\Gamma}}
\def\deltab{\bm{\delta}}            \def\Deltab{\bm{\Delta}}
\def\epsilonb{\bm{\epsilon}}
\def\varepsilonb{\bm{\varepsilon}}
\def\etab{\bm{\eta}}
\def\thetab{\bm{\theta}}            \def\Thetab{\bm{\Theta}}
\def\varthetab{\bm{\vartheta}}
\def\iotab{\bm{\iota}}
\def\kappab{\bm{\kappa}}
\def\varkappab{\bm{\vrkappa}}
\def\lambdab{\bm{\lambda}}          \def\Lambdab{\bm{\Lambda}}
\def\mub{\bm{\mu}}
\def\nub{\bm{\nu}}
\def\xib{\bm{\xi}}                  \def\Xib{\bm{\Xi}}
\def\pib{\bm{\pi}}                  \def\Pib{\bm{\Pi}}
\def\varpib{\bm{\varpi}}
\def\taub{\bm{\tau}}
\def\phib{\bm{\phi}}
\def\varpib{\bm{\varpi}}
\def\rhob{\bm{\rho}}
\def\varrhob{\bm{\varrho}}
\def\sigmab{\bm{\sigma}}            \def\Sigmab{\bm{\Sigma}}
\def\varsigmab{\bm{\varsigma}}
\def\phib{\bm{\phi}}                \def\Phib{\bm{\Phi}}
\def\varphib{\bm{\varphi}}
\def\chib{\bm{\chi}}
\def\psib{\bm{\psi}}                \def\Psib{\bm{\Psi}}
\def\omegab{\bm{\omega}}            \def\Omegab{\bm{\Omega}}
\def\taub{\bm{\tau}}
\def\uepsilonb{\bm{\uepsilon}}      \def\Uepsilonb{\bm{\Uepsilon}}
\def\zetab{\bm{\zeta}}
\def\gammabh{\wh{\gammab}}            \def\Gammabh{\wh{\Gammab}}
\def\deltabh{\wh{\deltab}}            \def\Deltabh{\wh{\Deltab}}
\def\epsilonbh{\wh{\epsilonb}}
\def\varepsilonbh{\wh{\varepsilonb}}
\def\etabh{\wh{\etab}}
\def\thetabh{\wh{\thetab}}            \def\Thetabh{\wh{\Thetab}}
\def\varthetabh{\wh{\varthetab}}
\def\iotabh{\wh{\iotab}}
\def\kappabh{\wh{\kappab}}
\def\varkappabh{\wh{\vrkappab}}
\def\lambdabh{\wh{\lambdab}}          \def\Lambdabh{\wh{\Lambdab}}
\def\mubh{\wh{\mub}}
\def\nubh{\wh{\nub}}
\def\xibh{\wh{\xib}}                  \def\Xibh{\wh{\Xib}}
\def\pibh{\wh{\pib}}                  \def\Pibh{\wh{\Pib}}
\def\varpibh{\wh{\varpib}}
\def\taubh{\wh{\taub}}
\def\phibh{\wh{\phib}}
\def\varpibh{\wh{\varpib}}
\def\rhobh{\wh{\rhob}}
\def\varrhobh{\wh{\varrhob}}
\def\sigmabh{\wh{\sigmab}}
\def\varsigmabh{\wh{\varsigmab}}
\def\phibh{\wh{\phib}}
\def\varphibh{\wh{\varphib}}
\def\chibh{\wh{\chib}}
\def\psibh{\wh{\psib}}                \def\Psibh{\wh{\Psib}}
\def\omegabh{\wh{\omegab}}            \def\Omegabh{\wh{\Omegab}}
\def\taubh{\wh{\taub}}
\def\uepsilonbh{\wh{\uepsilonb}}      \def\Uepsilonbh{\wh{\Uepsilonb}}
\def\zetabh{\wh{\zetab}}
\def\gammabt{\wt{\gammab}}            \def\Gammabt{\wt{\Gammab}}
\def\deltabt{\wt{\deltab}}            \def\Deltabt{\wt{\Deltab}}
\def\epsilonbt{\wt{\epsilonb}}
\def\varepsilonbt{\wt{\varepsilonb}}
\def\etabt{\wt{\etab}}
\def\thetabt{\wt{\thetab}}            \def\Thetabt{\wt{\Thetab}}
\def\varthetabt{\wt{\varthetab}}
\def\iotabt{\wt{\iotab}}
\def\kappabt{\wt{\kappab}}
\def\varkappabt{\wt{\vrkappab}}
\def\lambdabt{\wt{\lambdab}}          \def\Lambdabt{\wt{\Lambdab}}
\def\mubt{\wt{\mub}}
\def\nubt{\wt{\nub}}
\def\xibt{\wt{\xib}}                  \def\Xibt{\wt{\Xib}}
\def\pibt{\wt{\pib}}                  \def\Pibt{\wt{\Pib}}
\def\varpibt{\wt{\varpib}}
\def\taubt{\wt{\taub}}
\def\phibt{\wt{\phib}}
\def\varpibt{\wt{\varpib}}
\def\rhobt{\wt{\rhob}}
\def\varrhobt{\wt{\varrhob}}
\def\sigmabt{\wt{\sigmab}}
\def\varsigmabt{\wt{\varsigmab}}
\def\phibt{\wt{\phib}}
\def\varphibt{\wt{\varphib}}
\def\chibt{\wt{\chib}}
\def\psibt{\wt{\psib}}                \def\Psibt{\wt{\Psib}}
\def\omegabt{\wt{\omegab}}            \def\Omegabt{\wt{\Omegab}}
\def\taubt{\wt{\taub}}
\def\uepsilonbt{\wt{\uepsilonb}}      \def\Uepsilonbt{\wt{\Uepsilonb}}
\def\zetabt{\wt{\zetab}}

\def\d#1{\,\mbox{d} #1}
\def\pth#1{\left(#1\right)}
\def\acc#1{\left\{#1\right\}}
\def\cro#1{\left[#1\right]}
\def\rem#1{}

\def\zerob{{\bm 0}}
\def\oneb{{\bm 1}}

\def\intg{\int\kern-1.1em\int}
\def\expf#1{\exp\left[ {#1} \right]}

\def\bs{\bar{s}}
\def\us{\underline{s}}
\def\fmin{f_{\mbox{min}}}
\def\fmax{f_{\mbox{max}}}
\def\expf#1{\mbox{exp}\left\{#1\right\}}
\def\argmin#1#2{\mbox{arg}\min_{#1}\left\{#2\right\}}
\def\argmax#1#2{\mbox{arg}\max_{#1}\left\{#2\right\}}

\def\lra{\longrightarrow}
\def\fh{\widehat{f}}
\def\fbh{\widehat{\fb}}
\def\Rbh{\widehat{\Rb}}
\def\Sigmabh{\widehat{\Sigmab}}

\def\rbh{\widehat{\rb}}
\def\qbh{\widehat{\qb}}
\def\xbh{\widehat{\xb}}
\def\tbh{\widehat{\tb}}
\def\thetabhk{\widehat{\thetab}^{(k)}}
\def\thetabhkp{\widehat{\thetab}^{(k+1)}}

\def\disp{\displaystyle}
\def\vsm{\vspace*{-12pt}}
\def\hsm{\hspace*{-3em}}

\def\pmatrix#1{\left[\begin{matrix}#1\end{matrix}\right]}
\def\pmat#1{\left(\barr{c} #1 \earr\right)}
\def\pmata#1#2{\left(\barr{c} #1 \\ #2 \earr\right)}
\def\pmatb#1#2#3#4{\left(\barr{cc} #1 & #2 \\ #3 & #4 \earr\right)}

\def\th{\widehat{t}}
\def\xh{\widehat{x}}
\def\lambdah{\widehat{\lambda}}
\def\muh{\widehat{\mu}}
\def\sigmah{\widehat{\sigma}}
\def\rhoh{\widehat{\rho}}
\def\betah{\widehat{\beta}}
\def\alphah{\widehat{\alpha}}
\def\tbh{\widehat{\tb}}
\def\mubh{\widehat{\mub}}
\def\sigmabh{\widehat{\sigmab}}
\def\rhobh{\widehat{\rhob}}
\def\oneb{\mbox{\bf 1}}

\def\ie{i.e.,}
\def\iid{i.i.d.}
\def\apost{\emph{a posteriori }}
\def\aprio{\emph{a priori }}

\def\sumi{\sum_{i=1}^{M} }
\def\sumj{\sum_{j=1}^{N} }
\def\lambdah{\widehat{\lambda}}
\def\lambdabh{\widehat{\lambdab}}

\def\det#1{\mbox{det}\left(#1\right)}
\def\defined{\,\shortstack{$\triangle$\\ =}\,}
\def\sign{\hbox{sign}}

\def\ss{source separation }
\def\pdf{probability density function }
\def\mm{mixing matrix }
\def\sm{separating matrix }

\def\sbh{\widehat{\sb}}
\def\abh{\widehat{\ab}}
\def\Sbh{\widehat{\Sb}}
\def\Abh{\widehat{\Ab}}
\def\Bbh{\widehat{\Bb}}

\def\Lambdab{\bm{\Lambda}}
\def\xbh{\widehat{\xb}}

\def\pxta{p\left( \Xb | \Ab \right)}
\def\pxas{p\left( \Xb | \Ab, \Sb \right)}
\def\pasx{p\left( \Ab,\Sb | \Xb \right)}
\def\pax{p\left( \Ab | \Xb \right)}
\def\psx{p\left( \sb | \Xb \right)}
\def\pbx{p\left( \Bb | \Xb \right)}
\def\pxb{p\left( \Xb | \Bb \right)}

\def\ria{r_i\left( [\Ab^{-1} \xb]_i \right)}
\def\rib{r_i\left( [\Bb \xb]_i \right)}
\def\riat{r_i\left( [\Ab^{-1} \xb(t)]_i \right)}
\def\ribt{r_i\left( [\Bb \xb(t)]_i \right)}

\def\pia{p_i\left( [\Ab^{-1} \xb]_i \right)}
\def\pib{p_i\left( [\Bb \xb]_i \right)}
\def\piat{p_i\left( [\Ab^{-1} \xb(t)]_i \right)}
\def\pibt{p_i\left( [\Bb \xb(t)]_i \right)}

\def\zmat{\left[0\right]}
\def\lra{\longrightarrow}
\def\dpdx#1#2{\frac{\partial #1}{\partial #2}}
\def\diag#1{\mbox{diag}\left\{#1\right\}}

\def\KL#1#2{\mbox{KL}\left(#1~:~#2\right)}

\def\esp#1{\mbox{E}\left\{ #1 \right\}}
\def\espx#1#2{\mbox{E}_{#1}\left\{ #2 \right\}}
\def\Prob#1{\mbox{P}\left(#1\right)}
\def\var#1{\mbox{Var}\left(#1\right)}
\def\cov#1{\mbox{Cov}\left(#1\right)}
\def\corr#1{\mbox{Corr}\left(#1\right)}
\def\trace#1{\mbox{Tr}\left(#1\right)}
\def\rang#1{\mbox{rang}\left(#1\right)}

\def\sigmae{\sigma_{\epsilon}}
\def\Sigmae{\Sigmab_{\epsilon}}

\def\fraco#1#2{\left({#1}\right)\,/\,\left({#2}\right)}

\def\mjk{m_{jk}}
\def\pjk{p_{jk}}
\def\pjz{p_{j0}}
\def\pjkl{p_{jk}}
\def\sigmajk{\sigma_{jk}}
\def\Rbe{\Rb_{\epsilon}}

\def\zerob{{\bm 0}}
\def\oneb{{\bm 1}}
\def\intg{\int\kern-1.1em\int}
\def\expf#1{\exp\left[ {#1} \right]}

\def\gbu{\underline{\gb}}
\def\fbu{\underline{\fb}}
\def\zbu{\underline{\zb}}
\def\epsilonbu{\underline{\epsilonb}}
\def\thetabu{\underline{\thetab}}

\def\sigmae{{\sigma_{\epsilon}}}
\def\Sigmae{{\Sigmab_{\epsilonb}}}
\def\Sigmaf{{\Sigmab_{\fb}}}
\def\Sigmag{{\Sigmab_{\gb}}}

\def\fbuh{\widehat{\fbu}}
\def\zbuh{\widehat{\fbu}}
\def\thetabuh{\widehat{\thetabu}}

\def\fbh{\widehat{\fb}}
\def\Pbh{\widehat{\Pb}}
\def\Sigmabh{\widehat{\Sigmab}}
\def\Abh{\widehat{\Ab}}
\def\thetabh{\widehat{\thetab}}

\def\Rjk{{R_j}_k}
\def\fbik{{\fb_i}_k}
\def\fbjk{{\fb_j}_k}

\def\mik{{m_i}_k}
\def\sigmaik{{\sigma_i^2}_k}
\def\Sigmaik{{\Sigmab_i}_k}
\def\alphaik{{\alpha_i}_k}
\def\betaik{{\beta_i}_k}

\def\muik{{\mu_i}_k}
\def\vik{{v_i}_k}

\def\mjk{{m_j}_k}
\def\sigmajk{{\sigma_j^2}_k}
\def\Sigmajk{{\Sigmab_j}_k}
\def\alphajk{{\alpha_j}_k}
\def\betajk{{\beta_j}_k}

\def\mujk{{\mu_j}_k}
\def\vjk{{v_j}_k}
\def\njk{{n_j}_k}

\def\alphae{\alpha^{\epsilon}}
\def\betae{\beta^{\epsilon}}
\def\alphaez{\alpha^{\epsilon}_{0}}
\def\betaez{\beta^{\epsilon}_{0}}
\def\alphaei{\alpha^{\epsilon}_{i}}
\def\betaei{\beta^{\epsilon}_{i}}
\def\alphaeiz{\alpha^{\epsilon}_{i0}}
\def\betaeiz{\beta^{\epsilon}_{i0}}

\def\mbz{{{\mb}_z}}
\def\Sigmabz{{{\Sigmab}_z}}
\def\diag#1{\mbox{diag}\left[#1\right]}

\def\blue#1{{\color{blue}#1}}
\def\red#1{{\color{red}#1}}
\def\green#1{{\color{green}#1}}
\def\magenta#1{{\color{magenta}#1}}
\def\cyan#1{{\color{cyan}#1}}
\def\REM#1{}

\def\intd{\int\kern-.8em\int}

\def\cresta{
\thicklines\setlength{\unitlength}{.5mm}
\begin{picture}(120,120)
 \multiput(0,27)(0,20){5}{\line(1,0){80}}
 \multiput(0,27)(20,0){5}{\line(0,1){80}}
 \put(61,33){\red{$f_N$}}
 \put(0,92){\red{$f_1$}}
 \put(40,72){\red{$f_j$}}
 \put(88,50){\blue{$g_{i}$}}
 \put(38,120){\blue{$H_{ij}$}}
 \put(-5,115){\line(3,-2){90}}
 \put(42,85){\blue{\line(3,-2){20}}}
\end{picture}}

\def\tomoXaz{
\thicklines\setlength{\unitlength}{.96mm}
\begin{picture}(70,68)
  \put(15,35){\makebox(20,15){$\red{f(x,y)}$}}
  \put(-1,0){\axexy}
  \put(10,10){\objet}
  \put(5,5){\vector(1,1){55}}
  \put(55,55){\makebox(5,10){$r$}}
  \put(35,28){\makebox(5,10){$\phi$}}
  \put(60,8){\makebox(10,15){$\bullet$D}}
  \put(60,0){\makebox(15,10){$\blue{g(r,\phi)}$}}
  \put(10,55){\makebox(10,10){S$\bullet$}}
  \put(62,15){\line(-1,1){45}}
  \put(40,-5){\proj}
  \blue{\multiput(10,14)(5,0){9}{\line(0,1){40}}}
  \blue{\multiput(8,14)(0,5){9}{\line(1,0){40}}}
\end{picture}
}

\title{Non Gaussianity and Non Stationarity modeled through Hidden Variables and their use in ICA and Blind Source Separation}

\author{Ali Mohammad-Djafari}

\institute{Laboratoire des signaux et syst\`emes \\
UMR 08506 (CNRS, Sup\'elec, Univ Paris Sud 11)\\
Sup\'elec, Plateau de Moulon, 91192 Gif-sur-Yvette, France}

\begin{document}
\maketitle
\begin{abstract}
Modeling non Gaussian and non stationary signals and images has always been one of the most important part of signal and image processing methods. In this paper, first we propose a few new models, all based on using hidden variables for modeling either stationary but non Gaussian or Gaussian but non stationary or non Gaussian and non stationary signals and images. Then, we will see how to use these models in independent component analysis (ICA) or blind source separation (BSS). The computational aspects of the Bayesian estimation framework associated with these prior models are also discussed. 
\end{abstract}

\vspace{-24pt}
\section{Introduction}
In many signal and image processing methods, and in particular in ICA or in BSS, the first step is prior modeling of them. Here, we consider only the probabilistic modeling where the samples of a signal $\{f(t), t=1,\cdots,T\}$ are represented by a random vector $\fb$ for which we assign a probability law $p(\fb)$. The main problem is then to choose an expression for $p(\fb)$ to represent a particular family of signals or images. For example choosing a Gaussian expression for $p(\fb)=\Nc(\fb|\zerob,\Pb_0)$ with $\Pb_0$ an identity matrix will represent a stationary signal. In this case, we have $p(\fb)=\sum_t p(f(t))$ and the expression of $p(f(t))$ is Gaussian and does not depend on $t$. The main objective of this paper is to consider the cases where $p(\fb)$ is not Gaussian and/or is not separable and/or, if it is separable, $p(\fb)=\sum_t p_t(f(t))$ the expression of $p_t(f(t))$ depends on time $t$. In all these expressions, we can replace $t$ by $\rb$ representing the position index of a pixel for the case of images.  

One of the tools to model non Gaussianity is to use the mixture of probability laws, and in particular, the mixture of Gaussians: \\ 
\centerline{
\(
p(f(t))=\sum_{k=1} \alpha_k \Nc(f(t)|\mu_k,v_k)
\)}
\\ 
where $\thetab=\{(\alpha_k,\mu_k,v_k), k=1,\cdots,K\}$ are the parameters of the mixture and where $\sum_{k=1} \alpha_k=1$. When interpreting $\alpha_k=P(z(t)=k)$ with $z(t)$ a hidden variable, we can write 
\(
p(f(t)|z(t)=k)=\Nc(f(t)|\mu_k,v_k)
\) 
which gives the possibility to consider $z(t)$ as a classification label for the samples of the signal $f(t)$. But also, this gives the possibility to introduce non stationary in modeling $f(t)$ by letting $z(t)$ change in a given way with time. 

Another tool which also gives the possibility to introduce non gaussianity and non stationarity is to consider the parameters of the $p(\fb)$ to be random or change in time. One such example is:\\ 
\centerline{
\(
p(f(t)|v(t),\lambda)=\Nc(f(t)|0,2v(t)/\lambda ) 
\mbox{~~~and~~~}
p(v(t)|\lambda)=\Ic\Gc(v(t)|3/2,\lambda) 
\)
}
Here again, $p(f(t))$ is non Gaussian, and by letting $v(t)$ change with time, we can also obtain a non stationary signal. 

In this paper, we are exploring a few cases of such models, and in particular the mixture of Gaussians model with a hidden markovian model, for different applications. We consider, in particular, the case of ICA or BSS where these kind of models are used for the sources or for the components. 

The rest of this paper is organized as follows:  
In section II, a set of Gaussian/non Gaussian and/or stationary/non stationary models and their properties are presented. 
In Section III, we see how to use them as a prior law in a Bayesian framework, first in ICA and then in BSS. 
In Section IV, the Bayesian computational aspects related to the use of these models are discussed. 

\vspace{-12pt}
\section{Gaussian/Non Gaussian and stationary/Non stationary}

\vspace{-9pt}
\subsection{Gaussian and stationary models}
Let note the sample $f(t_j)=f_j$ and by $\fb=\{f_j, j=1,\cdots,T\}$ the whole samples and the 
Gaussian probability density function (pdf) $p(\fb)=\Nc(\fb|\fb_0,\Pb_0)$ with the mean 
$\fb_0$ and the covariance matrix $\Pb_0$. In a first step, we assume $\fb_0=\zerob$. 
Three particular cases are then of interest:
\bit
\item $\Pb_0=\sigma_f^2\Ib$. 
This is the case where $f_j$ are assumed centered, Gaussian and \iid:

\vspace{-18pt}
\beq
p(\fb)=\sum_j p(f_j) \propto \expf{-\frac{1}{2\sigma_f^2} \sum_j f_j^2}
\propto \expf{-\frac{1}{2\sigma_f^2} \|\fb\|^2}
\eeq
\vspace{-9pt}
\item $\Pb_0=\sigma_f^2\Cb\Cb^t$. This is the case where $f_j$ are assumed centered, Gaussian but correlated. the vector $\fb$ is then considered to be obtained by: 
\(
\fb=\Cb \xib 
\) 
with $\Cb$ corresponds to a moving average (MA) filtering and $p(\xib)=\Nc(\zerob,\sigma_f^2\Ib)$. 
In this case, we have:

\vspace{-18pt}
\beq
p(\fb)\propto \expf{-\frac{1}{2\sigma_f^2} \sum_j [\Cb\fb]_j^2}
\propto \expf{-\frac{1}{2\sigma_f^2} \|\Cb\fb\|^2}
\eeq
\item $\Pb_0=\sigma_f^2(\Db^t\Db)^{-1}$ with $\Db^t=(\Ib-\Ab)$. 
This is the case where $f_j$ are assumed centered, Gaussian and auto-regressive: 
\(
\fb=\Ab \fb + \xib 
\) 
with $\Ab$ a matrix obtained from the AR coefficients and $p(\xib)=\Nc(\zerob,\sigma_f^2\Ib)$. 
In this case, we have 

\vspace{-12pt}
\beq
p(\fb)\propto \expf{-\frac{1}{2\sigma_f^2} \|\Db\fb\|^2}
\eeq
A particular case of AR model is the first order Markov chain
\beq
p(f_j|\fb_{-j})=\Nc(f_{j-1},\sigma_f^2) \mbox{~~~with~~~} f_0=0
\eeq
with corresponding $\Ab$ and $\Db^t=\Ib-\Ab$ matrices \\ 

\centerline{\(
\Ab=\left[\barr{cccccc}
0 & 0 & . & . & 0\\ 1 & 0 & . & . & 0 \\ 0 & 1 & 0 & . & . \\ . &&&& . \\ 0 & . & . & 1 & 0
\earr\right], 
\Db^t=\left[\barr{cccccc}
1 & 0 & . & . & 0\\ -1 & 1 & . & . & 0 \\ 0 & -1 & 1 & . & . \\ . &&&& .\\ 0 & . & . & -1 & 1
\earr\right]
\) 
}
which give the possibility to write

\vspace{-12pt}
\beq
p(\fb)\propto \expf{-\frac{1}{2\sigma_f^2} \|\Db\fb\|^2} 
\propto \expf{-\frac{1}{2\sigma_f^2} \sum_j (f_j-f_{j-1})^2}
\eeq
\eit
These particular cases give us the possibility to extend the prior model to other more sophisticated non-Gaussian models which can be classified in three groups:
\bit
\item Separable: 
\vspace{-18pt}
\beq
p(\fb)=\prod_j p_j(f_j)\propto \expf{-\alpha \sum_j \phi_j(f_j)}
\eeq
where $\phi_j$ are positive valued functions. 
If $\phi_j=\phi, \forall j$, then the model is stationary. 
\item Markovian:
\vspace{-18pt}
\beq
p(\fb)=\prod_j p(f_j|f_{j-1})\propto \expf{-\alpha \sum_j \phi_j(f_j-f_{j-1})}
\eeq
where $\phi_j$ are positive valued functions called potential functions of the Markovian model.  
Again here, if $\phi_j=\phi, \forall j$, then we have a stationary (homogeneous) Markov model. 
\eit
Some examples of the $\phi$ expressions used in many applications are:  

\centerline{
\(
\phi(t)=\left\{ t^2;~ |t|^{\beta},1\le \beta\le 2;~ - t \ln t +1, t>0;~  \min(t^2,1);~ \frac{-1}{1+t^2} \right\}
\)
}

\vspace{-12pt}
\section{Modeling using hidden variables}
As we mentioned in introduction, hidden variables give the possibility to model NG and/or NS signals. We present here a few interesting cases. 

\smallskip\noindent\textbf{Energy modulated signals:~}
A simple model which can capture the energy or variance modulated signals is 
\cite{Idier01b}. 

\beq
p(f_j|v_j,\lambda)=\Nc(f_j|0,2v_j/\lambda) 
\quad\mbox{and}\quad 
p(v_j|\lambda)=\Gc(v_j|3/2,\lambda) 
\eeq
where $\Gc$ is the Gamma distribution. It is then easy to show the following relations:

\vspace{-15pt}
\beq
p(f_j,v_j|\lambda)
\propto\expf{-\lambda \left(\frac{f_j^2}{4 v_j^2}+v_j\right)} 
\eeq

\vspace{-15pt}
and 

\vspace{-15pt}
\beq
p(\fb,\vb|\lambda)
=\prod_j p(f_j,v_j|\lambda)\propto\expf{-\lambda \sum_j \left(\frac{f_j^2}{4 v_j^2}+v_j\right)}
\eeq
\noindent\textbf{Amplitude modulated signals:~}
To illustrate this with applications in telecommunication signal and image processing, we consider the case of a Gaussian signal modulated with a two level or binary signal.  
A simple model which can capture the variance modulated signal or images is
\beq
p(f_j|z_j,\lambda)=\Nc(z_j,2/\lambda) 
\eeq
\mbox{with} $z_j\in\{m_1=0,m_2=1\}$ and  
$P(z_j=m_k)=(1/2), k=1,\cdots,K=2$. 

It is then easy to show the following:
\beq
p(f_j|\lambda)
=\sum_{k=1}^K (1/2) \Nc(m_k,\sigma_k^2=2/\lambda)
\eeq
and $p(f_j|z_j,\lambda)\propto \expf{-\lambda (f_j-z_j)^2}$ and  $P(z_j=k|f_j,\lambda)\propto \expf{-\lambda (z_j-f_j)^2}$.

\smallskip\noindent\textbf{Mixture of Gaussians:~}
The previous model can be generalized to the general mixture of Gaussians. 
We then have the following relations:
\beq
\barr{lcl}
p(f_j|z_j=k,m_k,v_k)&=&\Nc(m_k,v_k=2/\lambda_k) 
\\ 
p(z_j=k)&=&\pi_k \quad z_j\in\{1,\cdots,K\}
\\
p(f_j|\pi_k,m_k,v_k)&=&\sum_{k=1}^K \pi_k \Nc(m_k,v_k) 
\earr 
\eeq
\vspace{-15pt}
and

\vspace{-9pt}
\beq
\barr{lcl}
p(\fb|\zb,\mb,\lambdab)
&\propto& \expf{-\sum_j\sum_k \lambda_k \delta(z_j-k) (f_j-m_k)^2}
\\
p(\zb|\fb,\mb,\lambdab,\bm{\pi})
&\propto& \expf{-\sum_j\sum_k[\lambda_k \delta(z_j-k) (f_j-m_k)^2+\ln \pi_k]}
\\
P(z_j=k|\fb,\mb,\lambdab)&\propto& \expf{-\lambda_k (f_j-m_k)^2+\ln \pi_k}
\earr 
\eeq
and 
\(
p(\fb,\zb|\sigmae^2,\mb,\lambdab) \propto \expf{-J(\fb,\zb)}
\) 
with
\beq
\barr{lcl}
J(\fb,\zb)
&=&\sum_k\sum_{\{j: z_j=k\}} \lambda_k (f_j-m_k)^2 
+\sum_k \ln(\pi_k) \sum_j \delta(z_j-m_k) 
\\
&=&\sum_k\lambda_k \|\fb_k-m_k\oneb\|^2 
+\sum_k n_k \ln(\pi_k) 
\earr 
\eeq
where $\mb=\{m_1,\cdots,m_K\}$,~~ $\lambdab=\{\lambda_1,\cdots,\lambda_K\}$,~~~  $\bm{\pi}=\{\pi_1,\cdots,\pi_K\}$, 
$n_k=\sum_j \delta(z_j-k)$ is the number of samples $f_j$ which are in the class $z_j=k$ and  $\fb_k=\{f_j : z_j=k\}$. For more details and applications of such modeling see \cite{snoussi04b,ichir06b}.

\smallskip\noindent\textbf{Mixture of Gauss-Markov model:~}
In the previous model, we assumed that the samples in each class are independent. 
Here, we extend this to a markovian model:

\vspace{-15pt}\beq
\barr{lcl}
p(f_j|z_j=k,z_{j-1}\not=k,f_{j-1},m_k,v_k)&=&\Nc(m_k,v_k)
\\ 
p(f_j|z_j=k,z_{j-1}=k,f_{j-1},m_k,v_k)&=&\Nc(f_{j-1},v_k) 
\\ 
P(z_j=k)&=&\pi_k \quad z_j\in\{1,\cdots,K\}
\earr 
\eeq
which can be written in a more compact way if we introduce 
$q_j=1-\delta(z_j-z_{j-1})$ by 
\beq
p(f_j|q_j,f_{j-1},m_k,v_k)=\Nc(q_j m_k +(1-q_j) f_{j-1},v_k)
\eeq
which results to:
\beq
\barr{l@{}c@{}l}
p(\fb|\zb,\mb,\lambdab)
&\propto& \expf{-\sum_j\sum_k\lambda_k \delta(z_j-k) [f_j-(q_j m_k+(1-q_j) f_{j-1})]^2}
\\
&\propto& \expf{-\sum_j\sum_k\lambda_k \delta(z_j-k) [(1-q_j)(f_j-f_{j-1})^2+q_j(f_j-m_k)^2]}
\earr 
\eeq
and 
\( 
p(\fb,\zb|\sigmae^2,\mb,\lambdab) \propto \expf{-J(\fb,\zb)}
\) 
with
\beq
\barr{l@{}cl}
J(\fb,\zb)
&=&\sum_j\sum_k \lambda_k \delta(z_j-k) [f_j-(q_j m_k+(1-q_j) f_{j-1})]^2 
+\sum_k n_k \ln(\pi_k)
\\ 
&=&\sum_j(1-q_j)(\tilde{f}_j-\tilde{f}_{j-1})^2 
+\sum_k n_k \ln(\pi_k)
\\ 
&=&\|\Qb\Db\fbt\|^2 
+\sum_k n_k \ln(\pi_k)
\earr 
\eeq
where $\tilde{f}_j=\lambda_{z_j}(f_j-m_{z_j})$, $\Db$ is the first order finite difference matrix and $\Qb$ is a matrix with $q_j$ as its diagonal elements. 

\REM{
A particular case of this model is of great interest: $m_k=0, \forall k$ and 
$\lambda_k=\lambda, \forall k$. Then, we have: 
\vspace{-15pt}
\beq
\barr{lcl}
p(f_j|q_j,f_{j-1},m_k,v_k)&=&\Nc((1-q_j) f_{j-1},v_k)
\\
p(\fb|\qb,\mb,\lambdab)&\propto& \expf{-\sum_j\lambda [f_j-(1-q_j) f_{j-1})]^2}
\\ 
&\propto&\expf{-\lambda \sum_j [(1-q_j) (f_j-f_{j-1})^2+q_j f_j^2]}
\earr 
\eeq
and 
\( 
p(\fb,\qb|\gb,\sigmae^2,\mb,\lambdab) \propto \expf{-J(\fb,\qb)}
\) 
\mbox{with}\quad 
\beq
J(\fb,\qb)
=\lambda \sum_j[(1-q_j) (f_j-f_{j-1})^2+q_j f_j^2]
+\sum_k n_k \ln\alpha_k
=\lambda \|\Qb\Db\fb\|^2+\sum_k n_k \ln\alpha_k
\eeq
where $n_k=\sum_j q_j$ is the number of discontinuities (length of the contours in the case of an image) $\alpha_k=P(q_j=1)$ and $1-\alpha_k=P(q_j=0)$. 
}
In all these mixture models, we assumed $z_j$ independent with $P(z_j=k)=\pi_k$. However, $z_j$ corresponds to the label of the sample $f_j$. It is then better to put a markovian structure on it to capture the fact that, in general, when the neighboring samples of $f_j$ have all the same label, then it must be more probable that this sample has the same label. This feature can be modeled via the Potts-Markov modeling of the classification labels $z_j$. In the next section, we use this model, and at the same time, we extend all the previous models to 2D case for applications in image processing and to MIMO applications. 

\vspace{-12pt}
\section{Mixture and Hidden Markov Models for images}
In image processing applications, the notions of contours and regions are very important. In the following, we note by $\rb=(x,y)$ the position of a pixel and by $f(\rb)$ its gray level or by $\fb(\rb)=\{f_1(\rb),\cdots,f_N(\rb)\}$ its color or spectral components. In classical RGB color representation $N=3$, but in hyperspectral imaging $N$ may be more than one hundred. When the observed data are also images we note them by $\gb(\rb)=\{g_i(\rb),\cdots,g_M(\rb)\}$. 

In ICA problems we have $\gb=\Ab\fb$ and in more general BSS problems, we have 
$\gb=\Ab\fb+\epsilonb$, where $\Ab$ is the mixing matrix. In ICA methods, one often assume  $\fbh=\Bb\gb$ where $\Bb$ is called separating matrix, which is ideally $\Bb=\Ab^{-1}$. 

For any image $f_j(\rb)$ we note by $q_j(\rb)$, a binary valued hidden variable, its contours and by $z_j(\rb)$, a discrete value hidden variable representing its region labels. We focus here on images with homogeneous regions and use the mixture models of the previous section with an additional Markov model for the hidden variable $z_j(\rb)$.  
\smallskip\noindent\textbf{Homogeneous regions modeling:~}
In general, any image $f_j(\rb), \rb\in\Rc$ is composed of a finite set $K_j$ of 
homogeneous regions $\Rjk$ with given labels $z_j(\rb)=k, k=1,\cdots,K_j$ such that 
$\Rjk=\{\rb ~:~ z_j(\rb)=k\}$, $\Rc_j=\cup_k \Rjk$ and the corresponding pixel values 
$\fbjk=\{f_j(\rb) ~:~ \rb\in\Rjk\}$ and $\fb_j=\cup_k \fbjk$. 
The Hidden Markov modeling (HMM) is a very general and efficient way to model appropriately 
such images. The main idea is to assume that all the pixel values 
${\fb_j}_k=\{f_j(\rb), \rb\in\Rjk\}$ of a homogeneous region $k$ follow a 
given probability law, for example a Gaussian $\Nc(\mjk\oneb,{\Sigmab_j}_k)$ where 
$\oneb$ is a generic vector of ones of the size ${n_j}_k$ 
the number of pixels in region $k$. 

In the following, we consider two cases:
\bit
\item The pixels in a given region are assumed iid:

\vspace*{-9pt}
\beq \label{hmm1}
p(f_j(\rb)|z_j(\rb)=k) = \Nc(\mjk,\sigmajk), \quad k=1,\cdots,K_j
\eeq
and thus
\vspace*{-9pt}
\beq \label{hmm1b}
 p(\fbjk|z_j(\rb)=k) = p(f_j(\rb), \rb\in\Rjk) = \Nc(\mjk\oneb,\sigmajk\Ib)
\eeq
This corresponds to the classical separable and mono-variate mixture models. 

\item The pixels in a given region are assumed to be locally dependent:
\beq \label{hmm2}
 p(\fbjk|z_j(\rb)=k) = p(f_j(\rb), \rb\in\Rjk) = \Nc(\mjk\oneb,{\Sigmab_j}_k)
\eeq
where $\Sigmajk$ is an appropriate covariance matrix. 
This corresponds to the classical separable but multivariate mixture models. 
\eit
In both cases, the pixels in different regions are assumed to be independent: 

\vspace{-12pt}
\beq \label{hmm2b}
p(\fb_j)=\prod_{k=1}^{K_j} p(\fbjk) = \prod_{k=1}^{K_j} \Nc(\mjk\oneb,{\Sigmab_j}_k).
\eeq

\smallskip\noindent\textbf{Modeling the labels:~}
Noting that all the models (\ref{hmm1}), (\ref{hmm1b}) and 
(\ref{hmm2}) are conditioned 
on the value of $z_j(\rb)=k$, they can be rewritten in the following general form

\vspace{-12pt}
\beq \label{hmm3}
p(\fbjk)=\sum_k P(z_j(\rb)=k) \; \Nc(\mjk,\Sigmajk) 
\eeq
where either $\Sigmajk$ is a diagonal matrix $\Sigmajk=\sigmajk\Ib$ or not. 
Now, we need also to model the vector variables $\zb_j=\{z_j(\rb), \rb\in\Rc\}$. 
Here also, we can consider two cases:
\bit
\item Independent Gaussian Mixture model (IGM), where 
$\{z_j(\rb), \rb\in\Rc\}$ are assumed 
to be independent and 
\vspace{-12pt}
\beq
P(z_j(\rb)=k)=p_k, \quad \mbox{with}\quad \sum_k p_k=1 \mbox{~~and~~} 
p(\zb_j)=\prod_k p_k
\eeq
\vspace{-12pt}
\item Contextual Gaussian Mixture model (CGM), 
where $\zb_j=\{z_j(\rb), \rb\in\Rc\}$ are assumed 
to be Markovian 
\vspace{-15pt}
\beq
p(\zb_j)\propto 
\expf{ \alpha \sum_{\rb\in\Rc} \sum_{\sb\in\Vc(\rb)} \delta (z_j(\rb)-z_j(\sb)) }
\eeq
which is the Potts Markov random field (PMRF). 
The parameter $\alpha$ controls the mean value of the regions' sizes. 
\eit

\vspace{-12pt}
\smallskip\noindent\textbf{Hyperparameters prior law:~}
The final point before obtaining an expression for the posterior probability law of all the unknowns, i.e, $p(\fbu,\thetab|\gbu)$ is to assign a prior probability law $p(\thetabu)$ to the hyperparameters $\thetabu$. Even if this point has been one of the main discussing points between Bayesian and classical statistical research community, and still there are many open problems, we choose here to use the conjugate priors for simplicity. The conjugate priors have at least two advantages:  
1) they can be considered as a particular family of a differential geometry based family of priors \cite{Snoussi04a}  and  
2) they are easy to use because the prior and the posterior probability laws stay in the same family. 
In our case, we need to assign prior probability laws to the means $\mjk$, to the variances $\sigmajk$ 
or to the covariance matrices $\Sigmajk$ and also to the covariance matrices of the noises $\epsilonb_i$ 
of the likelihood functions. 
The conjugate priors for the means $\mjk$ are in general the Gaussians 
$\Nc({\mjk}_0,{\sigmajk}_0)$, those of variances $\sigmajk$ are the inverse Gammas $\Ic\Gc(\alpha_0,\beta_0)$ and those for the covariance matrices $\Sigmajk$ are the inverse Wishart's $\Ic\Wc(\alpha_0,\Lambdab_0)$. 

\smallskip\noindent\textbf{Expressions of likelihood, prior and posterior laws:~}
We now have all the elements for writing the expressions of 
the posterior laws.  
We are going to summarizes them here: 
\bit
\item Likelihood: \qquad
\(
p(\gbu|\fbu,\thetabu)=\prod_{i=1}^M p(\gbu|\fbu,{\Sigmab_\epsilon}_i)
=\prod_{i=1}^M \Nc(\gbu-\Ab\fbu,{\Sigmab_\epsilon}_i)
\) \\ 
where we assumed that the noises $\epsilonb_i$ are independent, centered and Gaussian with covariance matrices ${\Sigmab_\epsilon}_i$ which, hereafter, are also assumed to be diagonal
${\Sigmab_{\epsilon}}_i={\sigma_{\epsilon}}_i^2 \Ib$. 

\item HMM for the images: \qquad
\(
p(\fbu|\zbu,\thetabu)=\prod_{j=1}^N p(\fb_j|\zb_j,{\mb}_j,{\Sigmab}_j)
\) \\ 
where we used $\zbu=\{\zb_j, j=1,\cdots,N\}$ and where we assumed that $\fb_j|\zb_j$ are 
independent. 

\item PMRF for the labels:  
\(
p(\zbu) \propto 
\prod_{j=1}^N \expf{ \alpha \sum_{\rb\in\Rc} \sum_{\sb\in\Vc(\rb)} \delta (z_j(\rb)-z_j(\sb)) }
\) \\ 
where we used the simplified notation $p(\zb_j)=P(Z_j(\rb)=z(\rb), \rb\in\Rc)$ and where we assumed $\{\zb_j, j=1,\cdots,N\}$ are independent. 

\item Conjugate priors for the hyperparameters:\\ 
\(
\barr{llllll}
p(\mjk)     =\Nc({\mjk}_0,{\sigmajk}_0),& 
p(\sigmajk) =\Ic\Gc(\alpha_{j0},\beta_{j0}),\\ 
p(\Sigmajk) =\Ic\Wc(\alpha_{j0},\Lambda_{j0}),& 
p({\sigmae}_i)=\Ic\Gc(\alpha_{i0},\beta_{i0}).
\earr
\)
\item Joint posterior law of $\fbu$, $\zbu$ and $\thetabu$
\vspace{-9pt}
\[ 
p(\fbu,\zbu,\thetabu|\gbu) \propto p(\gbu|\fbu,\thetab_1) \; p(\fbu|\zbu,\thetab_2) \; p(\zbu|\thetab_2) \; p(\thetabu)
\]
\eit

\vspace{-18pt}
\subsection{Bayesian estimators and computational methods}
The expression of this joint posterior law is, in general, known up to a normalisation factor. This means that, if we consider the Joint Maximum A Posteriori (JMAP) estimate
\vspace{-12pt}
\beq 
(\fbuh,\zbuh,\thetabuh)=\argmax{(\fbu,\zbu,\thetabu)}{p(\fbu,\zbu,\thetabu|\gbu)}
\eeq

\vspace{-9pt}
\noindent we need a global optimization algorithm, but if we consider the Minimum Mean Square Estimator (MMSE) or equivalently the Posterior Mean (PM) estimates, then we need to compute this factor which needs huge dimensional integrations. There are however three main approaches to do Bayesian computation:

\smallskip\noindent\textbf{Laplace approximation:~} 
When the posterior law is unimodale, it is reasonable to approximate it with an equivalent Gaussian which allows then to do all computations analytically. Unfortunately, very often, $p(\fbu,\zbu,\thetabu|\gbu)$ as a function of $\fbu$ only may be Gaussian, but as a function of $\zbu$ or $\thetabu$ is not. So, in general, this approximation method can not be used for all variables. 

\smallskip\noindent\textbf{Variational and mean field approximation:~} 
The main idea behind this approach is to approximate the joint posterior $p(\fbu,\zbu,\thetabu|\gbu)$ with another simpler distribution $q(\fbu,\zbu,\thetabu|\gbu)$ for which the computations can be done. A first step simpler distribution $q(\fbu,\zbu,\thetabu|\gbu)$ is a separable ones: 
\beq
q(\fbu,\zbu,\thetabu|\gbu)=q_1(\fbu) q_2(\zbu) q_3(\thetabu) 
\eeq
In this way, at least reduces the integration computations to the product of three separate ones. 
This process can again be applied to any of these three distributions, for example 
$q_1(\fbu)=\prod_j q_{1j}(\fb_j)$. With the Gaussian mixture modeling we proposed, $q_1(\fbu)$ can be chosen to be Gaussian, $q_2(\zbu)$ to be separated to two parts $q_{1B}(\zbu)$ and $q_{1W}(\zbu)$ where the pixels of the images are separated in two classes B and W as in a checker board. This is thanks the properties of the proposed Potts-Markov model with the four nearest neighborhood which gives the possibility to use 
$q_{1B}(\zbu)$ and $q_{1W}(\zbu)$ separately. For $q_3(\thetabu)$ very often we also choose a separable distribution which use the conjugate properties of the prior distributions. 

\smallskip\noindent\textbf{Markov Chain Monte Carlo (MCMC) methods:~} 
These methods give the possibility to explore the joint posterior law and compute the necessary posterior mean estimates. 
In our case, we propose the general MCMC Gibbs sampling algorithm to estimate $\fbu$, $\zbu$ and $\thetabu$ by first separating the unknowns in two sets $p(\fbu,\zbu|\thetabu,\gb)$ and  $p(\thetabu|\fbu,\zbu,\gbu)$. Then, we separate again the first set in two subsets $p(\fbu|\zbu,\thetabu,\gbu)$ and $p(\zbu|\thetabu,\gbu)$.  
Finally, when possible, using the separability along the channels, separate these two 
last terms in 
$p(\fb_j|\zb_j,\thetab_j,\gb_j)$ and $p(\zb_j|\thetab_j,\gb_j)$. 
The general scheme is then, using these expressions, to generates samples $\fbu^{(n)},\zbu^{(n)},\thetabu^{(n)}$ from the joint posterior law $p(\fbu,\zbu,\thetabu|\gbu)$   
and after the convergence of the Gibbs samplers, to compute their mean and to use them as the posterior estimates.

In this paper we are not going to detail these methods. However, we refer here to the application of these models in different area of signal and image processing and in particular in BSS  
\cite{Snoussi04a,bali06cc}.

\vspace{-12pt}
\section{Conclusion}
\vspace{-6pt}
In this paper, first we proposed a few new models for modeling either stationary but non Gaussian or Gaussian but non stationary or non Gaussian and non stationary signals and images. Then, we showed how to use these models in ICA or BSS. The computational aspects of the Bayesian estimation framework associated with these prior models are also discussed.    

\vspace{-12pt}
\def\sca#1{{\sc #1}}
\def\bibdir{/home/djafari/Tex/Inputs/bib/amd/}
\bibliographystyle{splncs}
\bibliography{bibenabr,revuedef,revueabr,baseAJ,baseKZ,gpipubli,\bibdir amd_art,\bibdir amd_ca,\bibdir amd_ci}

\end{document}